\documentstyle[12pt,epsf]{article}

\newlength{\dinwidth}
\newlength{\dinmargin}
\begin{document}

\newcommand{\be}{\begin{equation}}
\newcommand{\ee}{\end{equation}}
\newcommand{\ber}{\begin{eqnarray}}
\newcommand{\eer}{\end{eqnarray}}
\newcommand{\lp}{\left(}
\newcommand{\rp}{\right)}
\newcommand{\lk}{\left\{}
\newcommand{\rk}{\right\}}
\newcommand{\lc}{\left[}
\newcommand{\rc}{\right]}
\newcommand{\ls}{\alpha'}
\newcommand{\cm}{\hspace{1cm}}
\newcommand{\r}{{\bb R}}

\baselineskip18pt

\thispagestyle{empty}

\begin{flushright}
\begin{tabular}{l}
FFUOV-99/06
\end{tabular}
\end{flushright}

\vspace*{2cm}

{\vbox{\centerline{{\Large{\bf Non-perturbative Thermodynamics in
}}}}}

\vspace{0.8cm}

{\vbox{\centerline{{\Large{\bf
Matrix String Theory}}}}}

\vskip30pt

\centerline{Jes\'{u}s Puente Pe\~{n}alba
\footnote{E-mail address: jesus@string1.ciencias.uniovi.es}}

\vskip6pt
\centerline{{\it Depto. de F\'{\i}sica, Universidad de Oviedo}}
\centerline{{\it Avda. Calvo Sotelo 18}}
\centerline{{\it E-33007 Oviedo, Asturias, Spain}}

\vskip .25in

\vspace{1cm}
{\vbox{\centerline{{\bf ABSTRACT }}}}

   A study of the thermodynamics in IIA Matrix String Theory is presented.
The free string limit is calculated and seen to exactly reproduce the usual
result.  When energies are enough to excite non-perturbative objects like
D-particles and specially membranes, the situation changes because they add
a large number of degrees of freedom that do not appear at low energies.
There seems to be a negative specific heat (even in the Microcanonical
Ensemble) that moves the asymptotic temperature to zero. Besides, the
mechanism of interaction and attachment of open strings to D-particles and
D-membranes is analyzed.

\vspace*{24pt}

\setcounter{page}{0}
\setcounter{footnote}{0}

\newpage

\section{Introduction.}

   The Matrix Theory \cite{bfss} has supposed an important advance in
uncovering the global properties of M Theory or strongly coupled IIA String
Theory. It has answered several questions as well as offered a number of new
problems to be solved. Most problems are related to the fact that the model
is formulated in the light cone and it is precisely the light cone direction
the one that is taken to define the string coupling. This mixes the degrees
of freedom in such a way that calculations are simplified, but somehow lack
the easy interpretation they had in perturbative string theory. Objects are
not easily recognized and are, in fact, complicated combinations of the
asymptotic states of the Lorentz invariant theory. The study of its
thermodynamics was carried out in \cite{us3,sath,sath2,amb,zh} but it is
difficult to directly relate those calculations to the ones in usual
perturbative string theory.

   A partial remedy to this was set with the Matrix String Theory of
\cite{dvv,sfm} that put forward a non-perturbative formulation of String
Theory. The objects are still strings and the perturbative limit is not so
hard to find and analyze. In fact, the difference is in the way interactions
and non-perturbative objects appear. In the usual perturbative string theory
everything lies on the form of the world-sheet: its topological form
accounts for the different interaction terms while its singularities (fixed
points, boundary conditions...) are related to non-perturbative effects. In
the new approach, the world-sheet is always a circle and the dynamics is
defined by the Yang-Mills fields on it. Their interactions tell us about the
strings' interactions and their non-perturbative configurations are related
to D-branes.

   Previous works studied the general thermal scenario and the relation
among the `problems of Hagedorn' in the different matrix string theories
\cite{us4} and the perturbative free string limit \cite{gri}, that is also
obtained here from another perspective. The question I have attempted to
address in this article is what does this approach add to thermodynamics
when energies are beyond $g^{-1}$. I have analyzed how should
non-perturbative Yang-Mills configurations be included in the Free Energy
and the result is a more dynamical view of string and D-brane thermodynamics
in the whole energy range.

   The calculations support, in the appropriate limit, those done before (in
\cite{many}) and unveil the drastic consequences of including oscillating
membranes. The Hagedorn problem is worsened by the appearance of this even
more degenerate object and the final picture at very high energies seems to
be dominated by a single and very excited membrane at very low temperature.

\section{Review of the matrix strings from the finite temperature point of
view.}

   Let us start with the Hamiltonian of the Matrix Model compactified on
$S^1(light-cone)\times S^1$. It is

\be
H=\frac{1}{2}\frac{R_+}{ 2 \pi} \int_0^{2 \pi} d \sigma
tr \lk p_i^2+(DX_i)^2+ \theta^T D \theta+\frac{1}{R_9^2}\lp E^2 -\lc
X^i,X^j \rc^2 \rp + \frac{1}{R_9}\theta^T 
\gamma_i \lc X^i,\theta \rc \rk
\label{hamil}
\ee

   It is in $\alpha'(=l_p^3 R_9^{-1})$ units. Taking the limit of type IIA
Matrix String \cite{dvv} means equating $R_9=\sqrt{\alpha'} g_s$ and
assuming that $g_s$ is small. In that limit, configurations with
non-commuting matrices are very heavy and we can consider them decoupled or
include them as solitonic objects.

   The Hamiltonian of the perturbative objects is a free $U(1)^M$ gauge
theory that describes exactly the same fields as the Green-Schwarz
Hamiltonian plus a two-dimensional gauge field. This one is pure gauge and
can be taken to zero but for the solitonic configurations related to the
finite circle in which the theory is defined. We shall take those
configurations into account later.

   We shall now relate every quantum number of the $U(1)^M$ SYM to its
counterpart in String Theory. The light-cone variables are defined as
\ber
p^+=p_0+p_{11} \nonumber \\
p^-=p_0-p_{11}
\label{cone}
\eer

Let us remember that the light-cone Hamiltonian of strings is

\be
H=\frac{1}{2} (p^++p^-)
\label{ham1}
\ee
with
\be
p^-=\frac{p_i^2}{2 p^+}+\frac{1}{2 \ls p^+}\sum_n \lp \alpha^i_{-n} 
\alpha^i_{n}+ \tilde{\alpha}^i_{-n}  \tilde{\alpha}^i_{n} \rp
\label{ham2}
\ee
plus the fermionic part and with
\be
\sum_n \alpha^i_{-n} \alpha^i_{n}=N_{n_i} n_i
\ee
Here $p^+=\frac{M}{R}$ is discrete. The $n$ are the oscillation numbers.
Note that the second quantization of the string fields is already included
and is, of course, independent of the number $N$. This is the usual
treatment of multi-field systems in perturbative String Theory with a single
world-sheet.  In the SYM, the discrimination in different strings is related
to the `twisted states'. One can consider diagonal matrices in which the
elementary fields are not periodic, but, rather, have the following
property:
\ber
X_{i,i}(\sigma)&=&X_{i+1,i+1}(\sigma+2\pi) \nonumber \\
&\mbox{ ... }& \\
X_{i+k,i+k}(\sigma)&=&X_{i,i}(\sigma+2\pi) \nonumber
\eer

for a $k$-times-twisted state. This is allowed by the $Z^M$ symmetry that is
left even after taking the $g_s \rightarrow 0$ limit. The result is just the
same as if we considered that the world-sheet could have different lengths,
all multiples of the initial one. This is tantamount to rescaling the
prefactor of the Hamiltonian, that is the light-like momentum $p^+$. That is
why these twisted states represent the Kaluza-Klein momentum number of each
object in the eleventh dimension. They are equivalent in the $0+1$ SYM to
the bound states of the original D0-branes. The total Kaluza-Klein momentum
is therefore $M=\sum_k k M_k$, if $M_k$ is the number of $k$-times-twisted
strings. The sum in $M$, and particularly in each $M_k$ is the real second
quantization of strings because it allows to deal with several string
world-sheets at the same time. To find out the spectrum, each scalar field
has to be expanded in a Fourier series. The Hamiltonian of each term is a
harmonic oscillator with frequency
$\omega=2\pi n$ so their energy is
\be
H=\frac{M}{2R_+} + \frac{1}{2} \frac{R_+}{M \ls}\lc \sum_n n
\lp  N_n + \frac{1}{2}\rp+\sum_{\tilde{n}} \tilde{n} 
\lp \tilde{N}_{\tilde{n}} + \frac{1}{2}\rp \rc + 
\frac{R_+ p_i^2}{2 M}
\ee

   It exactly coincides with (\ref{ham1}) and (\ref{ham2}). We have focused
on the bosonic part. The fermionic one is a fermionic harmonic oscillator
that cancels the zero point energy of the bosonic one. In the original model
of D0-branes moving in a circle, these oscillator modes are mapped by the
flip and the Taylor construction \footnote{I do not explicitly call T-duality
to this construction because in this case it is not. From this M-theory point 
of view, $R_9$ and $R_+$ are just parameters and T- and S-dualities are just 
changes of names and variables.} \cite{taylor} to the modes of the open strings in which the
second end is stuck in the D-particle after having wound several times around 
the circle. $N$ is the total light-like momentum of the system.

   We shall now derive the known partition function of free strings from
this discrete light-cone Hamiltonian. It is
\be
Z=\exp \lc -\beta {\cal F}(\beta) \rc
\ee
where ${\cal F }$ is the free energy that can be calculated with
\be
-\beta {\cal F}(\beta)=Z_1=\sum_{\mbox{\tiny osc}}\sum_{M} \int d^8 
\vec{p} \exp \lk -\frac{\beta}{2} \lc \frac{M}{R_+} + \frac{R_+}{M} 
\lp \vec{p_T}^2 + m^2(n,N_n,{\tilde{n}},{\tilde{N}}_{\tilde{n}}) \rp \rc
\rk
\ee
if
\be
m^2(n,N_n,\tilde{n},\tilde{N}_{\tilde{n}})= N+\tilde{N}
\ee

  That is the total oscillator number. The integral over the continuous
momenta is Gaussian and can be performed. Besides, the sum over the
oscillator states can be decomposed in mass levels. This yields
\be
-\beta {\cal F}(\beta)=\sum_{N} \sum_{M} a_N \lp \frac{\beta R_+}{M} 
\rp^{4}
\exp \lc -\frac{\beta}{2} \lp \frac{M}{R_+} + \frac{R_+}{M}  N+R_+ \omega 
\rp \rc.
\label{lc}
\ee

 The coefficients $a_N$ can be related with arguments of number theory to
\be
a_N=\frac{1}{N!} \frac{d^N}{d x^N}\lc \theta_4 (0,x)^{-8} \rc_{x=0}
\label{coeff}
\ee

  This is the usual relation between thermodynamics and the string
non-thermal partition function. This is usually obtained after the
integration of the constraint that relates the left and right moving
oscillators. Here, it is got as a count of the number of ways a sum 
\be
\sum n N_n+\tilde{n} \tilde{N}_{\tilde{n}} 
\ee 
gives the same total $N$. The result is the same, no matter how different
they seem because of the properties of the Jacobi Theta functions.  The
constraint for the discrete light-cone states has been implemented as 
\be
N=\tilde{N}+\omega M 
\ee 
Expression (\ref{lc}) is the best we can get when
$R_+$ is finite. If we take the limit $R_+ \rightarrow \infty$, we can
obtain the usual form of the free energy. We have to substitute 
\ber
\frac{M}{R_+} & \longrightarrow & p_+ \nonumber \\ 
\sum_{M} & \longrightarrow & R_+ \int d p_+ 
\eer 

The states with $\omega\neq 0$
decouple when taking the infinite-$M$ limit. Also, in order to obtain a
proper time representation, we need to change variables 
\be p_+ \rightarrow \frac{N \beta}{t} 
\ee 
so that 
\be 
-\beta {\cal F}(\beta)=R_+
\beta^{-4}\sum_{N} \lp \frac{1}{N\beta} \rp^{-5} a_N \int \frac{dt}{t}
t^{-5} e^{-\frac{\beta^2 N}{4 t}} e^{-t} 
\ee

   In the right-hand side we can recognize the integral representation of the
Bessel function $K$, that we use to finally obtain

\be
-\beta {\cal F}(\beta)=R_+ \beta^{-4} \sum_{N} 
\lp \frac{2}{N\beta} \rp^{-5} a_N  K_5(\beta \sqrt{2 N})
\ee

   This has been obtained using classical statistics. Quantum statistics can
be added now as a series in $r$ substituting $\beta$ with $\beta r$.
Supersymmetry reduces the series to the odd terms:

\be
-\beta {\cal F}(\beta)=R_+ \beta^{-4} \sum_{N} \sum_{r,odd}
\frac{1}{r^{10}} \lp \frac{2}{N\beta} \rp^{-5} a_N  K_5(\beta r \sqrt{2N})
\label{closed}
\ee

   This is exactly what results following the $S$-representation or analog
model, that consists in separately summing the contributions of each of the
fields that appear at each mass level of the string. From this, using the
integral representation of the $K$ function and the coefficients
(\ref{coeff}), it is direct to obtain the usual modular invariant form. It
was got in a different way in \cite{gri}.

\section{The electric flux as a non-perturbative correction.}

   The last quantum number that we have got left to interpret is the
Kaluza-Klein momentum of the original D-particles related to the ninth
direction. It is mapped, in the matrix string limit, to the Ramond-Ramond
charge of the matrix D-particles \cite{dvv,bfm} (not the original, but the new
ones). These degrees of freedom, that were perturbative in the original matrix
theory, are now solitonic. The reason is that in every reduction of a YM action
to any number of dimensions, the number of degrees of freedom is $D-2$ except
in the reduction to $0+1$ (see \cite{us3}). In that case one more coordinate
appears in the free theory, which is essential to the interpretation of the
model as a theory in eleven dimensions. With the transformation to a $1+1$
theory that has been performed, that degree of freedom has vanished and only
left as a reminder a numerable set of non-equivalent topological sectors. Each
sector is characterized by the expectation value of momentum related to the
field $A_9$. The flux must be \be \Phi=\frac{1}{2\pi}\int d \sigma tr E=n,
\label{charge} \ee where $n$ is the Ramond-Ramond charge. That is the same as
to say $p_9=\frac{n}{R_9}$.

   Like in the previous section, I shall now give an interpretation to
every quantum number in the picture of String Theory. Remember that for a
general background that included both D0-branes and fundamental strings one
should separate the whole matrix into diagonal boxes, each containing one
twisted state. If the electric field related to the $U(1)^{M_i}$ diagonal
subgroup of a given box takes an expectation value, then that box describes a
D0-brane; else, it is a fundamental string.

The term of
the Hamiltonian that contains the electric field, is, in string units:
\be
H_E=\frac{R_+}{4 \pi} \int d\sigma \frac{E^2}{g_s^2}.
\ee 
If we consider simple, diagonal forms for $E$, the energy of a D-particle
state at rest is
\be                                                      
H_E=\frac{R_+}{2} \frac{n^2}{g_s^2}.
\ee

   As before, twisted states of these configurations represent particles
with different light-cone momenta.  Their energy is
\be                                                      
H_E=\frac{R_+}{2k}\frac{n^2}{g_s^2}=\frac{M_{D0}^2}{2p^+}.
\ee
   The $k$ is in the denominator because -remember- $E$ is not a field but a
momentum. Both numbers, $n$ and $k$, are completely independent. One can
have, for example $k \neq 1$ for $n=1$. That is got with a $k$-times-twisted
state with, for instance, $\Phi=k^{-1}$ for each matrix element. This is
possible because what must have an integer flux number is the $trace$ of the
field, not necessarily each matrix component.

   Each twisted state describes a D-particle. Bound states of D-particles in
String Theory cannot be seen like that, not even like twisted states in
this framework. Here, they are simply eigenstates of the ninth component of
the momentum with different eigenvalues or, in the dual system, field
configurations with different topological numbers (different $n$).

   The coordinate fields can also in this case be expanded in a Fourier
series. However, there is a subtlety here that has to be noticed. It is related
to the equations of motion of the scalar fields in the presence of the electric
field. To see it clearest, let me use the YM theory in the open space, whose
properties are more familiar. One equation is:

\be
\partial_\mu F^{\mu \nu}=0
\ee
which for the electric component and after a Fourier transform means that
\be
\vec{p} \vec{E}=0.
\ee

   This is the well known condition of transversality of the electromagnetic
waves. After the dimensional reduction, the only change is that both the
momentum and the electric field have only components in the remaining
directions. In our case, we only have one dimension and it is compactified.
The formula above implies that the electric field and the Kaluza-Klein momentum
exclude each other. In the previous section we studied the YM theory with
momentum (oscillator number) but without electric field. Now we are considering
the electric field so the theory is reduced to its zero Kaluza-Klein mode. 
This is the reason why D0-branes are pointlike even when they form twisted
states and the size of the matrix that contains them ($M$ in my notation) may be
taken to infinity. Another explanation to the absence of oscillators
(world-sheet KK momentum) can be obtained going back through the construction 
of W. Taylor 
to the original Matrix Model. Then the electric field is transformed into
the momentum along the ninth direction and the KK modes (oscillators) are the
 open strings
that stretch between the D-particles that form the one-dimensional lattice
that describes one D0-brane in a circle. The open strings are represented by
the massive gauge fields that appear when the symmetry is broken when the
positions of the D-particles acquire vacuum expectation value. Now, if we
consider the D-particle moving with a precise momentum, its conjugate
coordinate is completely unmeasurable and, therefore, it cannot have any
expectation value. This prevents the breaking of the gauge symmetry and the
appearance of any massive gauge field.

   This explains why D0-branes are pointlike in this approach. One important
consequence is that, for twisted states to be formed, the D-particles must
coincide at one point, unlike what happens with the closed strings. This is
not exactly the explanation given by \cite{dvv}. Their interpretation only
included D-particles with $M (p^+)= n (\mbox{D0 number})$, which is a strange relation between the
Dirichlet charge and the light-cone momentum. The string-like excitations
were eliminated from the spectrum just because their light-cone momentum was
set to zero in the large-$N$ limit.

   Before that, let us see how open strings can be attached to the D0-brane
in the perturbative limit. The mechanism, as pointed out in \cite{dvv} is
the same that causes interaction between different strings. When, at some
value of $\sigma$ the eigenvalues of all the spatial coordinates of the
D-particle coincide with those of a close string, some off-diagonal fields
get locally {\it nearly} massless. They are responsible for the attachment.
However, it seems to me that this case of D0-branes is a little bit subtler
than the strings' one. Firstly, whenever a closed string coincides at one
point with the position of the D-particle, there is not any exact
enhancement of symmetry. The reason is that the expectation value of $A_1$
is not proportional to the identity. The system is described by a matrix that
consists of two boxes. One is the D-particle and has an electric field. The other
is the matrix string. For simplicity, let us assume that both boxes have the
same size (both objects have the same light-cone momentum). In the D0-brane box
\be
\partial_0 A_1\propto E =\mbox{constant}
\ee
while in the string box the field can be set to $0$. Therefore
\be
A_1 \propto n t (\sigma_3+Id)
\ee
In the usual matrix string interaction, all fields coincide at one point and
there is an enhancement of symmetry. Here, the presence of the $\sigma_3$ field
slightly breaks that symmetry because off-diagonal fields do not become
completely massless. Let us compute its mass. 
The interesting term of the Hamiltonian is
\be
R_9[A^9,X^i]^2
\ee
We change
\be
(A^9, X^i)  \rightarrow R_9^{-1/2} (A^9, X^i)
\ee
for the fields to have appropriate dimensions and, taking into account that
\be
<(A^9)^2>\propto R_9^2= g^2 \ls
\ee
the corresponding mass of the mediating field is
\be
m^2 \propto g 
\ee 

   We are supposing that $g$ is very small, so the interchange of these very
light fields is permitted. To deduce from here that the coupling constant of
this interaction between the D-particle and the string is $g$, it should be
necessary to generalize the calculation in \cite{bonelli}. Another feature
that makes this attachment different from perturbative string interactions
is that it has more drastic consequences. The string loses half the
supersymmetry due to the existence of a privileged point that breaks the
invariance under reparameterizations. The point is distinguished in the
final world-sheet that combines both objects (and goes from $0$ to $4 \pi$)
as a discontinuity in the electric field, that is,
\ber
E(\sigma)=E  \cm \mbox{if} \cm  \sigma \in [0,2 \pi] \nonumber \\
E(\sigma)=0  \cm \mbox{if}  \cm \sigma \in [2 \pi,4 \pi] 
\label{conf}
\eer   
  The reason for that is that the twisted state crosses both boxes: the one
that represents the D-particle and that of the fundamental string.
  
   We have written before that in a the world-sheet of a single object the
electric field must be constant because any other configuration is
energetically very costly. The reason why (\ref{conf}) is possible is that
the process by which the electric field of some $U(1)$ extends itself to the
total group that includes both the string and the D-particle is clearly
non-perturbative. However, it is not completely forbidden: it is just the
absorption of a string and its light-cone momentum ($p_+$) by a D-particle
when both interact.

    In fact, the way the light-cone momentum is shared between the two
objects can be calculated. We have showed how the string is tied to the
D0-brane in all the transverse directions, but nothing has been said about
the longitudinal one. The difficulty is that the light-cone description we
have obliged the D-particle to be in the momentum representation. This
prevents the explicit appearance of a Dirichlet boundary condition. The only
reason for the string to be attached also in the ninth direction is
dynamical. We have to work with the covariant Hamiltonian one gets after
undoing the change of variables that took us to the light-cone. The free
part is
\be
H=\sqrt{M^2+\vec{p}^2_T+ (p^9_{\mbox{total}}-p^9_{\mbox{o.s.}})^2}+
\sqrt{m^2_{\mbox{o.s.}}+(p^9_{\mbox{o.s.}})^2}
\ee
where the subscript `o.s.' stands for open string, $M$ is the mass of the
D-particle and $m$ is the mass of the string. The minimum of the energy
occurs when
\be
\frac{p_{\mbox{o.s.}}}{p_{\mbox d0}}=\frac{m}{M}
\ee
This condition implies that the speeds of both objects coincide and
therefore, the dynamics of the open string is tied to that of the
D-particle. The link, however, is not very strong because the string can go
away with just the cost of its kinetic energy
\be
\frac{(p^9_{\mbox{o.s.}})^2}{2m}.
\ee

   This is due to the fact that when an open string is attached to a
D0-brane, its two ends are in the same point and so the string can close and
be emitted at no cost.

    It would be interesting to add these fields into the partition function
of (\ref{closed}). Once we throw away the part of the interaction and take
the $N\rightarrow \infty$ limit, the light-cone Hamiltonian is
\be
H_{l.c.}=\frac{p_+^{\mbox{d0}}}{2}+\frac{p_+^{\mbox{o.s.}}}{2}+
\frac{1}{2p_+^{\mbox{d0}}} \lc M^2+ (\vec{p}^{\mbox{d0}}_T)^2 \rc +
\frac{1}{2p_+^{\mbox{o.s.}}} \lc M^2+ (\vec{p}^{\mbox{o.s.}}_T)^2 \rc
\ee
with the constraint
\be
\vec{p}_T^{\mbox{d0}}=\frac{M}{m} \vec{p}^{\mbox{o.s.}}_T \cm \cm
p_+^{\mbox{d0}}=\frac{M}{m} p_+^{\mbox{o.s.}}
\ee

   It is straightforward to follow the calculation in the previous section
and to obtain the following proper-time representation of the Free Energy:
\be
-\beta F(\beta)=\sum_{\{ m, M \} } \beta \int \frac{dt}{t} t^{-5} \exp\lc 
-\frac{t}{2} \lp M+m \rp^2 - \frac{1}{2t} \beta^2 \rc 
\ee
We can now substitute the mass spectra
\ber
m^2=N \nonumber \\
M^2=\frac{n^2}{g^2}
\eer
where $N$ is the oscillator number of the attached string. This gives
\be
-\beta F(\beta)=\sum_{\{n, N \} } b_N  \beta \int \frac{dt}{t} t^{-5} 
\exp\lc - \frac{t}{2} \lp \frac{|n|}{g}+ \sqrt{N}  \rp^2 - \frac{1}{2t} 
\beta^2 \rc
\label{total}
\ee
 $b_N$ is the degeneracy associated to each mass level. This is the result
using classical statistics. The complete result has one more sum that
completes the exponential with $\beta$ to the usual Jacobi $\theta_3$ and
$\theta_2$ functions for bosons and fermions respectively.

   This does not look very familiar. If we want to see the open string's
Free Energy to emerge from this mess, we have to use the Born-Oppenheimer
approximation. It consists in classifying the degrees of freedom in fast and
slow ones. This approach was used in \cite{us3, amb} to study the dynamics
of D0-branes. This case is different because the light-cone direction is not
the one defined by $g$. This means that we have positive as well as negative
Ramond-Ramond charges and the excitations of them include the whole open
string spectrum, not just the massless sector. Anyway, the approximation is
valid as long as the mass of the D-particles is large. Here, the fast modes
are the open strings and the slow ones are the movements of the D-particles.
To see the open string partition function, we have to freeze the movements
of the D0-branes. We cannot do it with the proper time representation so we
have to go back to the Hamiltonian and set, there, all the momenta to zero
(or to a fixed value). The Hamiltonian is then
\be
H=\frac{|n|}{g}+ \sqrt{N}
\ee
   Index $n$ is fixed, we do not have to sum over it because we do not want
to allow the creation and annihilation of D-branes in our gas. To simplify,
consider just one D-brane. The Free Energy is simply
\be
-\beta F_n(\beta)=\sum_{r,odd} \sum_{N=0}^\infty b_N
e^{-\frac{|n|}{g}\beta -\sqrt{N} \beta r }
\ee

   There is not an $r$ multiplying the D-particle mass because that part is,
in fact, not thermalized at all; it plays the role of a cosmological
constant, it is just the energy of the vacuum.

   This representation of the Free Energy is not very recognizable so it is
better to put it in an integral form. The steps to follow are
\ber
-\beta F_n(\beta)=e^{-\frac{|n|}{g}\beta} \sum \exp(-\sqrt{N} \beta r)=
e^{-\frac{|n|}{g}\beta} \sum b_N \sqrt{\frac{2}{\pi}} (\beta \sqrt{N}r)^{1/2} 
K_{1/2}(\beta \sqrt{N}r)= \nonumber \\
= e^{-\frac{|n|}{g}\beta} \sum b_N \frac{1}{\sqrt{\pi}} \beta \int_0^\infty 
\frac{d t}{t}  t^{-1/2}  \exp (- \frac{\beta^2r^2}{2 t} - N t) = \nonumber \\ 
= e^{-\frac{|n|}{g}\beta} \frac{1}{\sqrt{\pi}} \beta \int_0^\infty 
\frac{d t}{t}  t^{-1/2}
\theta_2(0,\frac{\beta^2}{2 \pi t}) \frac{\theta_2^4(0,i t)}{\eta^{12}(i t)}
\eer

   The last expression is exactly the Free Energy of the open string theory
with Dirichlet boundary conditions in all directions, as obtained in the
S-representation. I have substituted the Laplace series
\be
\sum_N b_N e^{-N t}
\ee
with the appropriate Theta function. It is different from the one used for
closed strings because here the central charge breaks the supersymmetry and
the contribution of the fermions is changed: the supermultiplets are
shorter. I do not spend time with this because the procedure is known and
identical to that in the perturbative string theory.
 
   In this approach, the partition function of the slow modes represents a
count of the `possible vacuums' of the system. This is the way the existence
and positions of D-branes are seen in perturbative string theory. 

   Once we have seen that expression (\ref{total}) reproduces in the
appropriate limit the Free Energy of the attached strings, let us study the
structure when all the terms are taken into account. I am specially
interested in the modular properties.

   The first thing we have to do is to get rid of the square roots. We get
it in two steps; expanding the root and taking advantage of this trick:
\ber
\exp\lp -t \sqrt{N} \frac{|n|}{g} \rp =
\sqrt{\frac{2}{\pi}} \lp -t \sqrt{N} \frac{|n|}{g}
\rp^\frac{1}{2} K_{1/2}\lp t \sqrt{N} \frac{|n|}{g}
\rp= \nonumber \\
= \frac{t^{1/2}}{\sqrt{\pi}} \int \frac{d s}{s} s^{-1/2} \exp \lc
-\frac{1}{4s} \lp  t N  \frac{n^2}{g^2} \rp - s t \rc
\eer

   The cost is that we have added another integral. Altogether, the Free
Energy is
\ber
-\beta F(\beta)=\frac{\beta}{\pi} \sum \int \frac{d s}{s} \frac{d t}{t} 
s^{-1/2} t^{-11/2} b_N  \times \nonumber \\ \times
\exp \lc -\frac{1}{2t} \beta^2 (2m+1)^2- \frac{t}{2}\lp
\frac{n^2}{g^2} + N+ \frac{1}{2s}+ 2 s N \frac{n^2}{g^2}\rp \rc
\eer

We would like to take this mess and put some order, but, as far as I can
see, it is only partially possible. I have found two alternative ways to do
so:
\ber
-\beta F(\beta)=\frac{\beta}{\pi} \sum_{n\neq 0} \int \frac{d s}{s} \frac{d t}{t}
s^{-1/2} t^{-11/2} \theta_2\lp 0,i \frac{\beta^2}{\pi t}\rp
\times \nonumber \\ \times f \lc \lp 1+ 2s \frac{n^2}{g^2} \rp \frac{t}{2}
\rc \exp \lp -\frac{t}{2} \frac{n^2}{g^2} -\frac{t}{4s} \rp 
\eer

and

\ber
-\beta F(\beta)=\frac{\beta}{\pi} \sum \int \frac{d s}{s} 
\frac{d t}{t} s^{-1/2} t^{-11/2} b_N \theta_4\lp 0,i 
\frac{\beta^2}{\pi t}\rp
\times \nonumber \\ \times 
\lc \theta_3\lp 0,\frac{i t}{2 g^2} +\frac{i t s N}{g^2} \rp -1 \rc
\exp \lp - \frac{t}{2} N -\frac{t}{4s} \rp  
\eer

To simplify, I have defined $f$ to be the open string partition function 
\be
f(t)=\frac{\theta_2^4(0,i t)}{\eta^{12}(i t)}
\ee

   The introduction of D-branes as rigid non-perturbative corrections to
string thermodynamics has always yielded the same result: the critical
Hagedorn temperature does not change, although the properties of the
transition and the even its existence do depend on the background. The
reason for this stability is that, in fact and as far as D-branes are solely
static objects, the critical temperature only depends on the properties of
the world-sheet field theory. It is, therefore, an interesting topic to
decide if the addition of the dynamics of D0-brane changes anything. To know
that from the Canonical Ensemble, we have to calculate the behaviour of the
integrand of (\ref{total}) when $t \rightarrow 0$. To study the exponent
\be
\sum_{N,n} \exp \lp -\frac{t}{2}\frac{n^2}{g^2} - \frac{t}{2} N + 
\sqrt{8 \pi^2 N} -t \sqrt{N} \left| \frac{n}{g} \right| \rp
\ee
is enough. The coefficients $b_N$ have been substituted with their
asymptotic expression:
\be
b_N \sim N^{-9/4} e^{\sqrt{8} \pi \sqrt{N}}
\ee
It is possible to evaluate the sum in $N$ with a saddle point approximation.
The result of that is
\ber
\sum_{n} \exp \lp -\frac{t}{2}\frac{n^2}{g^2} + \frac{t}{2}\frac{n^2}{g^2}
-\sqrt{8 \pi^2}\left| \frac{n}{g} \right| +\frac{4 \pi^2}{4 t} \rp = 
\sum_{n}  \exp \lp -\sqrt{8 \pi^2}\left| \frac{n}{g} \right| + 
\frac{4 \pi^2 }{t} \rp
\eer

  The only dependence on $t$ is the last fraction that diverges at $t=0$.
The critical temperature is found when that term is compared to the first
mode of the thermal $\theta_4$ function:
\be
\frac{4 \pi^2 }{t}=\frac{\beta_c^2}{2 t} \Rightarrow \beta_c=\sqrt{8} \pi 
\ee

   This is the usual Hagedorn temperature both for (type II) closed and open
strings so the conclusion is that the dynamics of the D-particles with the
excitations we have considered do not change it. Remember that the spectrum
we have used is the free one, but that does not mean it is the complete one
even when $g=0$. The open strings that stretch between a couple of
D-particles have been considered in what precedes as part of the interaction
even when their contribution to the energy does not vanish at zero coupling.
The origin of this confusion is the non-commutativity of the D-brane
coordinate fields. When the off-diagonal terms are taken into account, the
separation into individual objects gets darker and the inclusion of their
contribution is a hard task. Maybe this states that we have not considered
are responsible for the apparent lack of modular invariance of the
expression obtained.

   In a very interesting work B. Sathiapalan \cite{sath} showed an aspect of
high temperature matrix theory that is related to this work. He used the $D=1+1$
field theory to deduce a `deconfinement-like' transition that he related to
Hagedorn. The energy of such transition would be non-perturbative so one may
question why the results and the methods of this section (and the following
ones) are so different. The answer is that the physical situations described are
different. The parameter that distinguishes both is the relative distance of the
objects or, if you want, the density of the gas. That separates two regions in
phase space according to whether the typical distance between objects is bigger
or smaller than $\ls$. If they are bigger, one must consider first the addition
of BPS non-perturbative objects like D-particles and D-membranes because their
energies are smaller than those of off-diagonal fields. This is what could be
called a dilute gas approximation and that is what is being carried out in this
work. If the density of the gas is such that typical distances are smaller than
the string length, then one should consider first the contribution of
off-diagonal modes. That is more difficult and up to now, only qualitative
information could be extracted in \cite{sath}.  D-objects carry
non-perturbative information but their geometrical nature makes them more
tractable.

\section{Membranes as D2-branes.}

   The flux that defines a membrane lying along the seventh and the eighth
directions is
\be
\Phi=\frac{i}{4 \pi}\int d\sigma tr < \lc X^7,X^8 \rc >=T_2 A .
\ee
where $A$ is the area of the membrane. It is necessary for that trace not to
be null that $N$ (the size of the matrices) should be infinite. If some
fields with that commutation relation take some expectation values, a
membrane is `created' from the vacuum. It is important not to confuse these
non-commuting fields with the centre-of-mass coordinates of the membrane
whose dynamics is described by the $N\times N$ identities that can be added
without changing the commutation relation. That is, the fields that give the
D2-brane charge are in the $SU(N)$ part. The expectation values of the
non-commuting matrix fields are related to the extension of the membrane in
each direction. Notice that neither the centre of mass nor the lengths are
affected by any uncertainty relation but the usual quantum one. Both
non-commuting fields can independently have any expectation values.

   As an example, we can choose the representation of the fields to be
equivalent to the position and momentum operators defined over a rectangle.
The correct normalization is such that their eigenvalues lie in the segment
$(3/2)^{1/3}[-1,1]$. In the usual quantum theory, the eigenvalues of the
coordinates and the momenta cannot be both continuous and bounded at the
same time; however, the reason has to do with the boundary conditions one
imposes to the eigenfunctions. There are not any boundary conditions in this
case, so that does not happen here. To fix notation let me write
\ber
\hat{X}^7=X^7 x \nonumber \\
\hat{X}^8=X^8 y  \\
\left[ x,y \right]=1. \nonumber
\label{mat}
\eer
Here $x$ and $y$ are operators (as matrices) that act over the Hilbert space
and $X^7$ and $X^8$ are numbers. However, as I have already written, they
are internal degrees of freedom and, as such, they are completely
independent of the position of the centre of mass. All excitations over this
background should be interpreted as the excitations of the membranes
(D2-branes in the perturbative limit).

   It is now our task to study some of the possible excitations of these
physical configurations. We shall begin with the quantization of precisely
the non-commuting fields so that we know what are their possible expectation
values, that is, the possible areas for the membrane. This is so because the
eigenvalues of the operators are the points in the target space that are
occupied by the membrane. The Hamiltonian that concerns them is
\be
H_{78}=p_7^2+p_8^2+\frac{1}{g_s^2} X_7^2 X_8^2,
\label{longitudinal}
\ee
where I have taken $\ls=1$ and I have been loose about constants. A first
approximation to the energy of the fundamental state is to consider that
$p x \simeq \hbar$ and minimize the Hamiltonian. The value of $X^7$ and
$X^8$ that are obtained are:
\be
X^7 \simeq X^8 \simeq g^{1/3} = l_p.
\ee

   This gives the very interesting result that the `smallest' possible
membrane is of the size of the eleven-dimensional Planck area. This also
increases the importance of the non-perturbative effects in string theory.
It is known that the energy density of D-objects is of order $g^{-1}$ in
string units; however, the result above means that the first membrane state
that is excited appears at an energy of order $g^{-1/3}$.

   These excitations that I have already discussed are longitudinal along
the directions of the membrane. The transversal ones can be described by
fields like
\be
X^6=f(x,y).
\ee
If we take one, for instance, to be proportional to $y$, its Hamiltonian,
with the correct normalizations would be
\be
H_6= \frac{1}{2 p^+} \lp p_6^2 + \frac{L_7^2}{g_s^2} X_6^2 \rp.
\ee
The coordinate is no longer free but rather, it oscillates in a harmonic
oscillator with frequency
\be
\omega=\frac{L_7}{g_s}.
\ee
That is precisely the one-dimensional tension of the membrane in the eighth
direction, that is, integrating out the other dimension. Explicitly:
\be
T_1^{(8)}=\int_0^{L^7} T_2=\int_0^{L^7} \frac{1}{g_s}=\frac{L_7}{g_s},
\ee
in $\ls$ units. We can now interpret these excitations as waves moving in
the eighth direction whose amplitude vibrates in the sixth. It is
straightforward to consider other excitations oscillating in other
directions and also moving in the seventh. The interpretation as waves is
quite clear because what the fluctuations actually do is to locally rotate
the membrane towards some direction and then leave it back where it was
lying.

   In fact, we can make a Fourier expansion of the fields moving in each
direction. We must impose appropriately chosen boundary conditions. If we
choose them to be Neumann, the expansion is made in terms of these fields:

\be
\Psi_n= \sqrt{2n} \sin \lp \pi n x  \rp
\label{transversal}
\ee
where $x$ is again the $N\times N$ matrix defined in (\ref{mat}). These
fields parameterize the waves moving in the seventh direction. Others do it
in the eighth and there are combinations that complete the Fourier
two-dimensional expansion of the membrane's vibrating modes. The mass these
fields acquire through the interaction with the background is
\be
E_n^2=N_n n \frac{L^7}{g_s}.
\ee
where $N_n$ is the excitation mode of the harmonic oscillator. Notice that,
in eleven-dimensional units,
\be
\frac{L^7}{g_s}=\frac{L^7}{l_p^3}
\ee
So that if we compactify the seventh direction and consider it the quantum
one (make a flip), the vibrations of the membrane in the other (eighth)
direction have the usual string spectrum proportional to the new $\ls(=l_p^3
(L^7)^{-1})$. Just like these, all possible vibrations of the membrane can
be constructed. These fields represent the second quantization of the
classical waves that deform the surface of the membrane. All of them are
excited only at non-perturbative energies so that we can conclude that in
the perturbative string limit the D-branes are completely rigid.

   Now, I shall explain how perturbative open strings can be formed. Let us
recall what gauge symmetry we have left after the non-commutative fields
have acquired their expectation values. As I have written before, the
matrices can be chosen to be coordinate and momentum (derivative) operators
acting over a one-dimensional space. This gives the correct commutation
relation and their eigenvalues give the eigenvalues of our matrices. There
are many $SU(\infty)$ matrices that obey that algebra. The different
possibilities are related to the different representations (coordinate,
momentum or mixed) of the basis that is used to span the Hilbert space.
Changing the representation is equivalent to changing the gauge. That is the
freedom we keep. This is related to the invariance under world-volume
reparameterizations inside the membrane. Notice that there is no statistical
$Z_N$ symmetry involved, which is coherent with the fact that we only have
one object.

\begin{figure}
%%Begin InstantTeX Picture
\let\picnaturalsize=N
\def\picsize{3in}
\def\picfilename{mems.ps}
%If you do not have the picture file add:
%\let\nopictures=Y
%to the beginning of the file.
\ifx\nopictures Y\else{\ifx\epsfloaded Y\else\input epsf \fi
\let\epsfloaded=Y
\centerline{\ifx\picnaturalsize N\epsfxsize \picsize\fi
\epsfbox{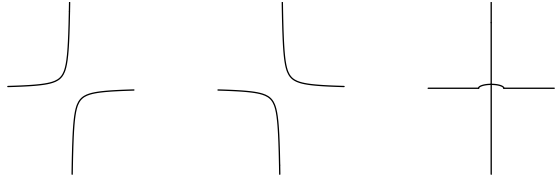}}}\fi
%%End InstantTex Picture
\caption{The three different parameterizations of two crossing membranes
(seen with little perspective).}
\label{mems}
\end{figure}

    Let us see some cases in which this symmetry is enhanced. Consider two
membranes that cross at one line. The case is identical if there is only one
membrane that crosses itself. One world-volume representation
would be
\ber
X^6_a(\sigma^1_a,\sigma^2_a)=\sigma^1_a  \nonumber   \\
X^6_b(\sigma^1_b,\sigma^2_b)=0  \nonumber     \\
X^7_a(\sigma^1_a,\sigma^2_a)=0   \nonumber  \\   
X^7_b(\sigma^1_b,\sigma^2_b)=\sigma^1_b     \\
X^8_a(\sigma^1_a,\sigma^2_a)=\sigma^2_a  \nonumber \\
X^8_b(\sigma^1_b,\sigma^2_b)=\sigma^2_b  \nonumber
\eer
where $a$ and $b$ index the membranes and $1$ and $2$ the directions. The
parameters $\sigma_{a,b}$ run inside $[-1,1]$, for example. The crossing
occurs at $\sigma^1_a=\sigma^1_b=0$. In the matrix case, there are two
appropriate representations of $\hat{X}^8$ as a momentum of the other
coordinates at that line. They are related to the different ways how the
momentum (derivative) can flow across the membrane. Figure \ref{mems}
explains it. Mathematically, one can define $X^8$ as a derivative (at
$\sigma^1=0$) in all these ways:
\ber
\left. \partial_a f \right|_{\sigma=0}=\lim_{\Delta \sigma\rightarrow 0}
\frac{f(\sigma_a+\Delta \sigma_{a/b})-f(\sigma_a)}{\Delta\sigma_{a/b}} 
\nonumber \\
\left. \partial_b f \right|_{\sigma=0}=\lim_{\Delta \sigma\rightarrow 0}
\frac{f(\sigma_b+\Delta \sigma_{b/a})-f(\sigma_b)}{\Delta\sigma_{b/a}} 
\eer

   There are three different possibilities that correspond to the three ways
one can separate two crossing planes into two surfaces. These
representations are, in the matrix language, gauge equivalent matrices. This
gauge enhancement can be seen as responsible for perturbative
splitting-and-joining interactions between membranes.

   A similar argument can be used to explain the attachment of the open
strings to the D-brane.

   Consider, then, a matrix that can be `diagonalized' into two boxes: one is
the membrane and the other is a usual matrix string state. There can be an
enhancement of symmetry if at least at one particular value of $\sigma$ (the
original world-sheet) the coordinate of the string coincides with one of
the target space points occupied by the D-brane. That means that their
transverse coordinates must coincide and that the coordinates of the string
which are paralel to the membrane must be equal to one of the eigenvalues of
each non-commuting matrix.

   In that case, it is possible to exchange each matrix (seen as a
derivative operator acting over the Hilbert space in the representation of
the operator of the other direction)
\be
\left. \partial f \right|_i= f(i+1)-f(i-1)
\ee  
with another like this:
\ber
\left. \partial f \right|_i= \left( f(i+1)-f(i-1) \right) \nonumber \\
+\delta_{\kappa,i+1} \times \left( f(\kappa)-f(i+1) \right)
-\delta_{\kappa,i-1} \times \left( f(\kappa)-f(i-1) \right)
\eer
if $\kappa$ is the index that marks the field of the string. We have not
cared much about the normalizations. This change is, in fact, equivalent to
interchanging the matrix indices $i$ and $\kappa$ so that it is a symmetry
for both matrices, $\hat{X}^7$ and $\hat{X}^8$ (this is important). This
symmetry has to exist in any representation of the matrices (any
parameterization), in particular, both where one of them is diagonal. I
point this out to remark the necessity for both eigenvalues of the position
of the string to coincide with a point inside the membrane. This is a
locally (in $\sigma$) conserved gauge symmetry. In this case the enhancement
is to a $Z_2$ symmetry (instead of $SU(2)$ from a $Z_2$, like in the case of
closed strings).

\begin{figure}
%%Begin InstantTeX Picture
\let\picnaturalsize=N
\def\picsize{3in}
\def\picfilename{twist.ps}
%If you do not have the picture file add:
%\let\nopictures=Y
%to the beginning of the file.
\ifx\nopictures Y\else{\ifx\epsfloaded Y\else\input epsf \fi
\let\epsfloaded=Y
\centerline{\ifx\picnaturalsize N\epsfxsize \picsize\fi
\epsfbox{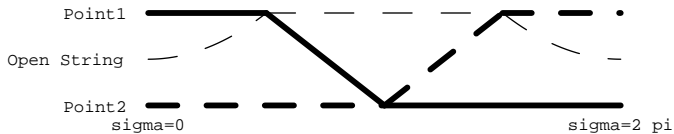}}}\fi
%%End InstantTex Picture
\caption{Twisted state of an open string with two membrane points. As
$\sigma$ runs, the fields in the non-commuting $SU(N)$ part cannot oscillate
or move (horizontal lines) because their excitations are non-perturbative
but the open string field can go from one point to the other.}
\label{twist}
\end{figure}

   This symmetry allows perturbative pointlike interactions between strings
and D2-branes. Besides it allows the creation of a kind of `twisted states'
that are related to the attached open strings and the boundary closed string
states. Figure (\ref{twist}) is a graphic explanation. Wherever two lines
coincide a $Z_2$ twist is performed and one field takes the role of the
other. There can be two special points (or just one) that mark the two ends
of the open string. As can be seen in the figure, the field that comes out
of the membrane after the twist to substitute the string (straight lines
with slope) is the way the momentum flows back between the two ends of the
string. As a final result of the twist the fields that represent the two
positions inside the membrane (thick lines) are interchanged. This is
possible thanks to the same symmetry that has been described above. In a
membrane state without any open string, the interchange of the value of two
diagonal fields is not permitted because it would change the commutator with
the matrix in the other direction. In this case, the twist has not only
swapped the eigenvalues, but also the momentum-like matrix in such a way
that the total action is part of the complete reparameterization group.
Somehow, this is the recovery of part of the usual Weyl group that appears
in perturbative Matrix String Theory. Moreover, one can consider two
membranes so that the two special points are one in each of them. Then the
swap of indices strongly reminds the fact that stretching open strings are
represented in the usual SYM D-brane action as off-diagonal fields with the
two indices ($X^{ij}$ if $i$ and $j$ are the indices of the branes) and
therefore those that mediate their interchange.

   When an open string is part of a twisted state, its only movement can be
to continuously change its endpoints. For that it is essential that $N$ is
infinite so that the spectrum of eigenvalues of the matrices in continuous.
The centre of mass is tied to the membrane and the only way out it to join
the two endpoints so that the twisted state can break without loss of
energy.

    Once a state like that has been excited, interactions with other
external strings are easier to understand because away from the fixed
points, the strings behave just like ordinary matrix strings and interact as
usual.

\section{Membranes with different charges and different geometries.}

   A membrane with opposite charge to the one studied in the previous
section would be, for instance
\ber
\hat{X}^7=X^7 x \nonumber \\
\hat{X}^8=- X^8 y \nonumber \\
\left[ x,y \right]=1,
\eer
so that
\be
\left[ \hat{X}^7,\hat{X}^8 \right]=- X^7 X^8 \times Id.
\ee

They are related by a change in the orientation of the base that
parameterizes them. A representation of the fields was
\be
\hat{X^8}=\frac{\partial}{\partial X^7}
\ee
for the previous case while for the anti-brane it is
\be                                                    
\hat{X^8}=-\frac{\partial}{\partial X^7}.
\ee

   The conclusion from this is that it is possible to associate to every
membrane the vector product of the vectors of the basis over its surface. It
represents its orientation in the sense that it goes inwards in one face of
the membrane and outwards in the other. If the vectors of two membranes have
the same direction, they have the same charge; else, they are oppositely
charged.

   This is also true looking directly at the supersymmetry algebra of the
membranes as BPS states, but the form it adopts is different. The
Ramond-Ramond charge can be seen as a central charge of the algebra in the
form of a constant anti-symmetric tensor with two indices. It is a tensor
that transforms covariantly under Lorentz transformations. One can, in
particular, consider as before a membrane lying in the seventh and the
eighth directions. Its charge has the form
\be
Z^{78}=-Z^{87}=C.
\ee 
It is very easy to see that after a $180^o$ rotation, the charge changes to
\be
Z'^{78}=-Z'^{87}=-C.
\ee

   This will have interesting consequences when discussing membranes of
different geometries, which is done in the following. It also explains, in a
very geometrical way why the coupling to the Ramond-Ramond forms is opposite
in each case. Let us consider, for simplicity, the coupling of the massless
level of the closed string to the D2-brane. It only couples to the graviton
and to the Ramond-Ramond three-form. The coupling to the graviton involves
only the energy-stress tensor and, therefore, it is universal and
independent of the charge. However, the form couples through the integral
\be
\int d x_7 d x_8 A^{780}.
\ee

   Changing the orientation of the base that spans the plane means changing
the sign of the product $d x_7 d x_8$ or, equivalently, interchanging the
limits of one of the integrals. That is why
\be
\int d x'_7 d x'_8 A^{780}=-\int d x_7 d x_8 A^{780}
\ee 
and the coupling to the form has the opposite sign.

   Once we have studied plane membranes, it is not difficult to construct
the fields we need to describe any other membrane. The algorithm is: take a
parameterization of the surface and then equate two locally perpendicular
coordinates to any representation of the two non-commuting matrices ($x$ and
$y$) that have been used in the previous section. An easy example is a
cylinder. The fields we need are

\ber
\hat{\phi}=\phi \times x \nonumber \\
\hat{X^8}=X^8 \times y
\eer

with

\be
[x,y]=1.
\ee

Then

\ber
\hat{X^8}=X^8  \nonumber \\
\hat{X^7}=\sin \hat{\phi}  \nonumber \\
\hat{X^6}=\cos \hat{\phi},
\eer
with an appropriate normalization. These three fields describe the
cylindrical membrane. We have needed one more parameter than in the plane
case because it extends itself in three dimensions. The proportion between
the expectation values of $X^6$ and $X^7$ is related to the excentricity of
the ellipse that appears cutting the cylinder perpendicularly to its height.
It is very interesting to see that even though the energy of this background
corresponds to that of a membrane with that form and surface ($T_2 \times
A$), the total D2-charge is zero because the trace of both commutators
($[\hat{X}^7,\hat{X}^8]$ and $[\hat{X}^6,\hat{X}^8]$) vanish and thus all
the supersymmetry is exact and this state cannot be considered BPS. What
happens is that, {\cal locally}, we can define a conserved charge, but when
one integrates over the surface (sums the trace up) the charge of each point
is cancelled against the diametrically opposed one. This is directly related
to the rotational properties of the charge that have been described a little
above. When the surface curves around itself, the orientation of the inner
face continuously turns so that the points that stand in front of each other
possess opposite charges. In that sense, a self-interaction can occur such
that a closed string gets out of the membrane at one point and is reabsorbed
at the diametrically opposed one. This string interchange has the same
properties as that of the brane-antibrane problem. In particular, if the
cylinder is small enough, some open-string states become tachyonic and the
system unstable. The conclusion is that this cylindrical membranes are only
semi-stable, and the larger their radii are, the more stable. If the radius
of the cylinder becomes of order $\ls$, non-perturbative interactions
involving off-diagonal fields that transport D2-charge annihilate the
membrane and the supersymmetry is recovered in all space.

   This analysis can be directly generalized to the torus, the sphere and
any other closed surface. It is important to remark that these are curved
surfaces embedded in a flat space-time. If the membrane is toroidal but is
wrapping around a toroidal space, this is not valid. In that case,
`opposite' points are not related by a $180^o$ rotation, but rather by a
$\pi R$ translation. This, of course, does not change the orientation and
therefore the total charge does not vanish.

   If we want to describe those wrapped membranes, we can perform a
T-duality of the compactified theory. Then the definition of the central
charge is
\be
\Phi=\frac{i}{4 \pi} \int d\sigma d\sigma_7 d\sigma_8 tr 
< \lc A^7,A^8 \rc >=T_2 A 
\ee
and the potential term
\be
V=\frac{1}{g_s^2}\int d\sigma d\sigma_7 d\sigma_8 tr 
\lc A^7,A^8 \rc^2.
\ee

   It is clear that a proper choice of the constant fields (proportional to
$x$ and $y$, for example) one can obtain configurations with non-zero
charge.

   Another interesting situation is a background where two or more membranes
of the same charge coincide. There, we should be able to see an enhancement
of the symmetry to a local (in the world-volume) $SU(N)$ gauge symmetry.
This happens indeed. The system is described by a matrix that can be put in
the form of two independent boxes, each representing one membrane. When all
the expectation values of all the identity matrices that represent the
centres of mass of the membranes are identical, these two fields
\be
\lp \begin{array}{cc} 0 & I \\ I & 0  \end{array}
\rp \cm \mbox{and} \cm \lp
\begin{array}{cc} 0 & i I \\ -i I & 0  \end{array} \rp
\ee
become massless and generate a {\cal global} $SU(N)$ ($SU(2)$ in this case)
symmetry. The local symmetry is related to the freedom the boundary closed
strings (open strings in the other picture) have to form `twisted states' of
the kind described in the previous section. The states that connect the two
membranes are now massless and are responsible for the $SU(N)$ symmetry
restoration at every point in the world-volume. An even better way of
understanding it is to notice that, indeed, the possible choice of two
operators as derivative of the dual ones is local, that is, it can be done
for every point in the world-volume.

\section{Thermodynamical effects of membranes.}

   As has already been commented, the study of the effects of static,
non-dynamical D-branes in string thermodynamics has been studied at length
in several articles \cite{many}. What I would like to add is a glimpse of
what happens to the systems when the thermodynamical energies are high
enough to create pairs of membranes and anti-membranes. In the traditional
scenarios, the branes were stable because they possess a conserved charge
and so cannot decay; however, when energies are of order $g^{-1}$ a finite
density of non-perturbative objects should appear even if the total charge
is zero. In a certain sense, their enormous mass only acts as a chemical
potential that obstructs their nucleation. This affects the system in a
different manner if we use the Canonical or Microcanonical Ensemble and if
we suppose the total energy to be finite or we deal with densities.

\begin{figure}
%%Begin InstantTeX Picture
\let\picnaturalsize=N
\def\picsize{3in}
\def\picfilename{tempe.ps}
%If you do not have the picture file add:
%\let\nopictures=Y
%to the beginning of the file.
\ifx\nopictures Y\else{\ifx\epsfloaded Y\else\input epsf \fi
\let\epsfloaded=Y
\centerline{\ifx\picnaturalsize N\epsfxsize \picsize\fi
\epsfbox{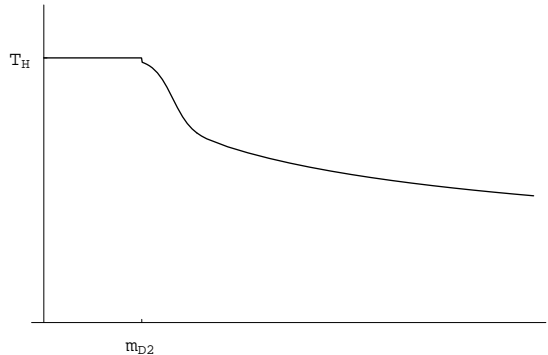}}}\fi
%%End InstantTex Picture
\caption{Temperature of a IIA string gas at non-perturbative energies.}
\label{tempe}
\end{figure}

   The picture is clearer if we suppose we are in the Microcanonical
Ensemble and the total energy is finite. In that case, the low energy
behaviour is `protected' from non-perturbative effects (that means:
unaffected by them) thanks to their mass. Only when the energies are high
enough should we consider them. We have seen that the spectrum of membranes
includes the quantization of the classical modes. The number of these modes
grows very fast with the energy but it is not possible (rather: we do not
know of any way) to calculate their degeneracy exactly; nevertheless, there
is a computation in \cite{ealvarez} that supplies an asymptotic form for the
density of mass states. The formula is
\be
\Omega(m)=\exp (m^{4/3})
\ee   
This should approximately count the degeneracy of the modes in formulas
(\ref{longitudinal}) and (\ref{transversal}) and others alike. It only
considers toroidal membranes but it is enough to see the asymptotic
behaviour. They are the non-pertubative excitations that change the form of
the membrane and come as a quantization of its classical oscillation modes.
If we assume that the temperature at those energies is very near Hagedorn,
the density of states we should use is
\be
\Omega(E)={\cal H}\lp E-\frac{1}{g^{1/3}}\rp \exp (E^{4/3})+ {\cal H}
\lp \frac{1}{g^{1/3}}-E \rp e^{\beta_H E}
\ee
 The term in the right corresponds to the gas of strings at the Hagedorn
temperature. It gives a behaviour shown in figure \ref{tempe}. I have
supposed that the membranes begin having their asymptotic behaviour soon
after their first modes are excited. The temperature can be seen to be
decreasing so that the specific heat is negative. The asymptotic temperature
for very high energies is zero. This is a consequence of the fact that
membranes as well as higher dimensional objects have senseless
thermodynamics. Nevertheless, the sense is recovered in this case thanks to
the large `chemical potential' that is the mass. The negative specific heat
is the result of a continuous series of relativistic mass thresholds being
opened. It is exactly the same phenomenon as Hagedorn, but with the
asymptotic temperature equalling zero. The huge mass degeneracy makes it
much more entropic to accumulate energy in the form of mass than in the form
of kinetic energy (temperature).

   The situation when one takes the thermodynamical limit ($V \rightarrow
\infty$ and $E \rightarrow \infty$) is worse because any individual object
can survey the whole energy range. In this case equipartition is broken from
the very beginning and, in principle, one membrane would absorb everything
around it until the universe were completely frozen. This picture, however,
has to be changed a little. Since we are dealing with the membrane as a
single object, the situation becomes similar to the appearance of black
holes at finite temperature. Black holes are always much more degenerate
than anything around them and classically one should expect them to grow and
grow until nothing else exists. Fortunately, this is changed when the first
quantum corrections are introduced: the black hole can and does reach
equilibrium with a gas of massless fields, its own Hawking radiation. This
happens because of two things: firstly, the degeneracy of the black hole,
large as it is, is not infinite and it always leaves some space for the rest
of the objects; and secondly, interaction makes these objects (the membrane
as well as the black hole) have a finite probability to emit energy in the
form of strings. Another source of instability for membranes is that we
assume as a natural fact that the total D2-charge of the universe is zero.
This means that pairs of membranes can decay into a very large number of
strings.

   All this phenomena (emition, pair annihilation and scattering) heat the
gas that gets a finite temperature when the energy density is also finite.
Presumably this temperature should be lower than Hagedorn and go down to
zero when the density tends to infinity.

   A more serious problem is trying to see the perspective that offers the
Canonical Ensemble. The simple attempt to calculate the Free Energy is
hopeless:
\be
\beta F(\beta)=\int d E \Omega(E) e^{-\beta E}=\int d E e^{-\beta 
E} e^{\beta E^{4/3}}=\infty
\ee

   It has not got any sense at any temperature. This is, of course, due to
the fact that the behaviour of the system is by no means regular at any
temperature.

\section{Speculations about higher dimensional objects and further comments.}

   After an intense study of many years \cite{hag,us12} on the problem of
Hagedorn, many conclusions have been achieved but still many doubts remain.
The physical picture has evolved but, although some non-perturbative effects
were added \cite{many} the problem has remained more or less the same
because only the dynamics of the strings was studied. The emergence of a
fully eleven-dimensional M-theory with unwrapped free membranes seems to
change things beyond non-perturbative energies. Moreover, after we have seen
the drastic effects of the oscillating modes of membranes, one could wonder
what could happen when higher dimensional objects like five-branes enter the
scene. Any calculation is impossible at this moment because we do not know
the precise spectrum and can only guess whether all the classical modes of a
five-brane can be excited on the M-five-brane or not. The same happens with
the nine-brane. If we naively accept that they do, the result at extremely
high temperatures is the preeminence of the single nine-brane, that is, the
whole space-time oscillating in high frequency modes. As non-perturbative
deformations of space-time, black holes should have an important role to
play.

   Another thing that I want to explain is why branes with different
dimensions behave so differently when, in the light of T-duality, they are
basically the same kind of objects. To understand this better, we look at
the construction of T-duality in \cite{taylor}. A D0-brane moving in a
space compactified in a circle becomes equivalent to a wrapped D1-brane
moving in the dual space.  The clearest image is an intermediate one that
consists in noticing that a D0-brane in a circle is equivalent to a linear
lattice of them in an open space with a constant separation. This means
that, in fact, the wrapped D-string dynamics is equivalent to that of an
infinite number of D-particles. Following the argument the dynamics of a
wrapped D-membrane is reproduced by a bidimensional rectangular lattice of
D-particles. However, the D-string or the D-membrane that come out of this
description are rigid and must be wrapping a torus. The D-particle cannot
reproduce the oscillations of the higher dimensional objects. To put it
short: the dynamics of lower dimensional D-branes are fully contained in
that of the higher dimensional ones, but not vice versa.

\section{Conclusions.}

   Throughout this work, I have tried to incorporate the advances in the
knowledge of non-perturbative corrections to String Theory that has yielded
the Matrix String Theory. The picture is more dynamical and clearer than had
been the previous images of Dirichlet branes. For objects with two
dimensions or less, the construction is powerful enough to allow quite a
complete treatment of the whole spectrum.

   After the first, qualitative approach of the global thermal consequences
made in \cite{us4}, a more quantitative calculation was in need. After
reproducing the usual perturbative string limit, I have introduced the
effects of free D-particles. This includes the bound states among
themselves, with closed strings and, also, the open strings whose two ends
are tied to the same D-particle. The expression obtained is not explicitly
invariant under modular transformations, possibly because not all the open
string states were included; however it has a critical temperature that
precisely equals the Hagedorn one.

   Regarding the membranes, the complexity of their spectrum makes it much
harder to obtain exact expressions. However it is possible to find out the
symmetry that fixes the open strings to the D2-brane and that is responsible
for their perturbative interaction. Apart from this usual perturbative
states, I have found the fields that represent the quantization of the
classical oscillations of the membrane as an extended object with a given
tension. The inclusion of these states changes drastically the
thermodynamics at non-perturbative energies and beyond. The canonical
description seems imposible while in the Microcanonical Ensemble, the system
there is dominated by a single large membrane extremely excited that tends
to absorb everything that comes in its way. This cools the universe so much
that the asymptotic temperature is zero.

   Maybe the most graphic way to understand this cooling is in a more or
less cosmological manner. Take the universe to be filled with an enormous
energy and that we slowly decrease it. At the beginning we have a cold gas
in thermal equilibrium with the excited membranes. The number of membranes
tends to two because the total D2-charge must be zero. As we extract energy
from the system, it gets hotter because the membranes are more and more
unstable and tend to annihilate between themselves. Eventually they
completely dissolve in the string gas that heats up to the Hagedorn
temperature. In that moment the picture is the one drawn in \cite{us12}
with a fat string in equilibrium with the surrounding sea.      

  One of the reasons for this strange behaviour is the absence of modular
invariance, that is lost when the eleventh dimension is discovered. Maybe
there is another symmetry that involves higher dimensional objects
(five-branes) that might correct this in part.

\section*{Acknowledgements.}

   I have to thank M. Laucelli Meana and M. A. R. Osorio for discussions on
this topic (and many others).

\end{document}